\documentclass[12pt]{spieman}  
\usepackage{amsmath,amsfonts,amssymb}
\usepackage{graphicx}
\usepackage{setspace}
\usepackage{tocloft}
\usepackage{bm}
\usepackage{xspace}
\usepackage{xcolor}
\usepackage{dutchcal}
\usepackage{courier}
\usepackage{rotating}
\usepackage{hyperref}

\newcommand\numberthis{\addtocounter{equation}{1}\tag{\theequation}}
\newcommand{\IRSSquare}{IRS$^2$\xspace}
\newcommand{\bnm}[1]{\bm{\mathbf{#1}}}

\newcommand{\eg}{{\it e.g.}\xspace}
\newcommand{\vs}{{\it vs.}\xspace}
\newcommand{\etc}{{\it etc.}\xspace}
\newcommand{\etal}{{\it et al.}\xspace}

\newcommand{\HgCdTeS}{HgCdTe\xspace}

\xspace
\newcommand{\mbb}[1]{\mathbb{#1}}
\newcommand{\msc}[1]{\mathcal{#1}}

\title{Simple Improved Reference Subtraction\\ for H4RG, H2RG, and H1RG Near-infrared Array Detectors}

\author[a,*]{Bernard J. Rauscher}
\author[b]{Dale J. Fixsen}
\author[a]{Gregory Mosby Jr.}
\affil[a]{Observational Cosmology Laboratory, NASA Goddard Space Flight Center, 8800 Greenbelt Rd., Greenbelt, USA, 20771}
\affil[b]{CRESST/UMd/GSFC, Greenbelt, USA, 20771}
\affil[ ]{\small Published in: JATIS 8(2), 028002 (2022), doi: 10.1117/1.JATIS.8.2.028002}

\cftpagenumbersoff{figure}
\cftpagenumbersoff{table} 
\begin{document} 
\maketitle

\begin{abstract}
Teledyne's H4RG, H2RG, and H1RG near-infrared array detectors provide reference pixels embedded in their data streams. Although they do not respond to light, the reference pixels electronically mimic normal pixels and track correlated read noise. In this paper, we describe how the reference pixels can be used with linear algebra and training data to optimally reduce correlated read noise. Simple Improved Reference Subtraction (SIRS)  works with common detector clocking patterns and, when applicable, relies only on post-processing existing data so long as the reference pixels are available. The resulting reference correction is optimal, in a least squares sense, when the embedded reference pixels are the only references and the reference columns on the left and right are treated as two reference streams. We demonstrate SIRS using H4RG ground test data from the Nancy Grace Roman Space Telescope Project. The Julia language SIRS software is freely available for download from the NASA GitHub.\cite{Download} The package includes a python-3 ``backend'' that can be used to apply SIRS corrections if a SIRS calibration file has been provided by the instrument builders.
\end{abstract}


{\noindent \footnotesize\textbf{*}Bernard J. Rauscher,  \linkable{Bernard.J.Rauscher@nasa.gov} }

\section{Introduction}\label{sec:intro}

We describe a simple and effective post-processing technique for reducing the correlated read noise in many near-infrared (NIR) astronomy images. Simple Improved Reference Subtraction (SIRS) uses the reference columns that are built into all Teledyne H4RG, H2RG, and H1RG (collectively ``HxRG'') arrays to efficiently suppress correlated read noise. Compared to current practice, SIRS automatically provides reference corrections that are optimal in a least squares sense without any tuning.

SIRS builds on the same mathematical foundation as the James Webb Space Telescope (JWST) Near Infrared Spectrograph's (NIRSpec) Improved Reference Sampling and Subtraction (\IRSSquare; pronounced ``IRS-square''),\cite{Moseley2010a,Rauscher2017} although it does not require the specialized NIRSpec clocking pattern. Instead, SIRS relies on traditional clocking patterns and post processing. It brings some of the benefits of \IRSSquare to all JWST NIR instruments for observers who write their own calibration software. SIRS is also applicable to archival data so long as the reference pixels are still available. Like \IRSSquare; SIRS makes reference corrections that are linear, deterministic, and optimal using least squares as the figure of merit.

\subsection{SIRS in the Context of Other Work}\label{sec:context}

Most reference correction is done in the time domain. Areas of reference pixels are averaged or a median is computed, and the result is subtracted from photosensitive pixels. The JWST pipeline implementation is typical.\cite{JDox} The reference rows and columns are used differently, and the reference rows correction tends to be more widely used.

Because they are sampled only at the beginnings and ends of frames, but appear in all outputs, the reference pixels in rows are used to remove constant offsets between outputs. This is done by robustly computing a spatial average or median of the reference pixels in rows in an output and subtracting it from every pixel in that output. By treating even and odd columns separately, one can also begin to remove some of the alternating column pattern noise (ACN) that is sometimes seen in H2RG detector systems.

We have not seen ACN in H4RG data. But, when it is present in H2RG images, it appears as faint vertical bars in alternate columns.  In Fourier space, ACN appears as a feature at the Nyquist frequency with a $1/f$-like shoulder on the low-frequency side. Physically, we believe that the alternating column pattern at the Nyquist frequency likely carries much of the same $1/f$ that appears at low frequency, but up-converted to the Nyquist frequency. We refer the interested reader to Rauscher (2015)~\cite{Rauscher2015} for more information on ACN.

When the reference columns are used, JWST's implementation is typical. A reference pixel signal is calculated using a running median of the reference columns on the left and right. These are optionally smoothed using a kernel having tunable width, multiplied by a tunable gain parameter, and subtracted from the photosensitive pixels. Schlawin~\etal (2020)\cite{Schlawin2020} provides a good recent introduction to this and other advanced time domain techniques in the context of JWST NIRCam.

SIRS is different. Like JWST's \IRSSquare, it is built on the observation that the read noise is approximately covariance stationary over a frame readout time. Fourier space therefore provides a much less correlated representation than time space. SIRS then seeks the best linear projection of the references onto the data using least squares as the figure of merit.

With careful tuning, it has been learned that the JWST pipeline can approach the cosmetic performance of SIRS in rejecting $1/f$ banding. But, SIRS still enjoys quantitative advantages, and moreover these advantages are realized without the need for any tuning and one knows that the resulting correction is optimal if least squares is accepted as the figure of merit.

\subsection{When Should SIRS be Considered?}\label{sec:help-me}

SIRS's main benefit is reducing correlated low frequency read noise. To quickly determine whether SIRS might help in a particular situation, the key diagnostic is visible banding in slope images (see Fig~\ref{fig:img-comp}; we defer detailed discussion of this figure to Section~\ref{sec:test}). If banding is visible, and if the typical band has a height that is more than two lines, then SIRS should help. A ``slope image'' is the result of least squares fitting straight lines to up-the-ramp sampled IR array data.

More quantitatively, SIRS works when there is temporal correlation in the pixel time series that is longer than the time required to read two lines. The factor of two is needed because the reference columns must critically sample correlated noise to remove it. In Fourier space, this becomes $f_\textrm{noise} < f_\textrm{Ny,line}$, where
\begin{equation}
f_\textrm{Ny,line} = \frac{1}{2 t_\textrm{line}}.\label{eq:nyquist1}
\end{equation}
Here, $t_{line}$ is the time required to read one line including any overheads for starting the next line.

Returning to Fig~\ref{fig:img-comp}: (1) there is visible banding and (2) the height of a typical band appears to be a few lines.This is a situation where SIRS should help.

\subsection{Paper Outline}\label{sec:outline}

The rest of this paper is structured as follows.

Section~\ref{sec:sci} presents the scientific rationale for why reducing correlated noise is important.  This includes a discussion of why correlated noise is particularly important to Roman.

In Section~\ref{sec:tech-intro}, we briefly describe the technical features of HxRG detectors that are most important for this paper. These include the reference pixel layout and a description of the conventional clocking pattern that was used. Teledyne's user manuals describe several clocking options. This section explains which options were used here.

We then describe the theory in Section~\ref{sec:theory}. This includes describing how least squares is used for optimization and presenting the key equations. This section includes a new derivation that can be easily extended to incorporate additional references as they become available.

SIRS's development was motivated and enabled by the large data sets that the Goddard Detector Characterization Laboratory (DCL) is producing for Roman. In Section~\ref{sec:test}, we use these data to show how SIRS works in practice.

Finally, Section~\ref{sec:future} describes future work. This includes modifying SIRS to work with the flight clocking pattern that will be very different from what was used here.

We close with a summary.

\section{Science Case}\label{sec:sci}

Mostly, SIRS reduces correlated noise. Although there is some reduction in the (uncorrelated) variance, since it is only a few percent it is barely significant in practice. However, correlated noise is more pernicious. There are many scientific studies that could be adversely affected by the types of correlated noise that SIRS removes.

At the most basic level, nearly all NIR astronomy uses more than one pixel to make measurements. When this is done, whatever the measurement, one must consider the covariance within the set of pixels that is used. By definition, uncorrelated noise falls along the diagonal of the covariance matrix and one can follow the familiar rule of adding uncertainties in quadrature. But, this breaks down when correlation is present and simply adding the uncertainties in quadrature risks badly under-estimating the true uncertainty.
 
The pernicious effects go far beyond propagation of errors however. Roman's weak lensing survey provides an example. It uses the gravitational distortion of faint field galaxy shapes to probe the growth of structure in the universe.  The correlations addressed here cause banding, and tend to raise or lower groups of lines in the detector. This in turn tends to exaggerate the extent of galactic images in the rapid scan direction.  For galaxies at the edge of visibility this can be a significant distortion. Even for galaxies that are well above the noise, the correlated noise can still complicate measuring the shear field. Although weak lensing groups are expert in these things, we believe it is nevertheless better to just remove the correlated noise if possible before the careful scientific analysis starts.
 
A second example is provided by correlations between galaxies. A small amount of correlated noise can enhance galaxies just enough to push them over the detection threshold.  Since the noise lines appear at distinct intervals on the detector, they will appear at some angle on the sky. That in turn will manifest at some preferred spatial angle on the sky.

Another important class of science observation that would benefit from improved reference correction is intensity mapping. Instead of just detecting individual objects, \eg galaxies, a number of pixels in a compact region are co-added to improve sensitivity. The science return is derived from looking at the statistical properties of such regions. An example is infrared background anisotropy, which will figure prominently in the Euclid and SPHEREx missions. The data reduction may amount to taking angular power spectra over extended regions and may be directly impacted by the structure shown in slope images derived with standard reference pixel subtraction.

There can be many other examples where the subtle effects of correlated noise that appear at or even below the random noise level, but by combining many images or integrating over many sources can be enhanced, just like the signal, to become the dominant source of noise. For $2^\textrm{nd}$ order measurements like correlations, galaxy distortions, \etc, correlation not only generates noise --it can generate bias.

\section{Technical Description of HxRG Detectors}\label{sec:tech-intro}

Teledyne's HxRG detector arrays are among the most widely used astronomy sensors today. All HxRGs have a photosensitive mercury-cadmium-telluride (\HgCdTeS) detector layer bonded to a silicon readout integrated circuit (ROIC). The ROIC routes control voltages and clocks to pixels and multiplexes the many individual pixel outputs to a much smaller number of detector outputs. For easy reference, Table~\ref{tab:hxrg-layout} summarizes some of the key parameters that relate to SIRS. Readers who desire a more thorough introduction to HxRGs may wish to see Mosby~\etal\cite{Mosby2020} and references therein.

\begin{table}[h]
\caption{HxRG Formats}
\centering
\begin{tabular}{ccccc}
\hline\hline
Parameter& Unit& H4RG& H2RG& H1RG\\\hline
Pixels readout&
	pixels&
	$4096\times 4096$&
	$2048\times 2048$&
	$1024\times 1024$\\
\HgCdTeS pixels&
	pixels&
	$4088\times 4088$&
	$2040\times 2040$&
	$1016\times 1016$\\
Number of outputs&
	\#&
	1, 4, 16, 32 or 64&
	1, 4, or 32&
	1, 2, or 16\\
\hline 
\end{tabular}
\label{tab:hxrg-layout}
\end{table}%

Every instrument will be different and the appearance of visible banding in slope images remains the best indicator for when SIRS might be beneficial. Broadly speaking though, we would expect SIRS to work best with short frame readout times. At the $10^5~\textrm{pixels}~\textrm{s}^{-1}~\textrm{output}^{-1}$ readout rate that is typically used, we would expect SIRS to potentially benefit H4RG and H2RG systems using $\geq32$ outputs and H1RG systems using 16 outputs.

Although JWST's H2RG based Near Infrared Camera (NIRCam) and Near-Infrared Imager and Slitless Spectrograph (NIRISS) do not satisfy these conditions, our preliminary analysis of JWST ground test data suggests that they nevertheless benefit. Although the reduction in variance and visible banding is small compared to carefully tuning all features that are built into the pipeline, we believe that further study may show significant statistical benefits when using JWST near its detection limits.

\subsection{Reference Pixels}

\begin{figure}[h]
    \centering
    \includegraphics[width=.8\columnwidth]{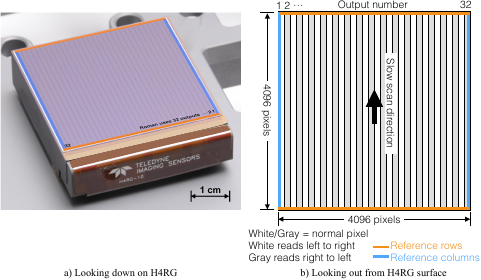}
    \caption{We developed SIRS for use with Roman ground test data. a) Roman's Wide Field Instrument (WFI) will fly eighteen $4096\times 4096$~pixel Teledyne H4RGs. b) The WFI configuration uses 32 outputs as shown here. All HxRGs embed a 4-pixel wide border of reference pixels in the data streams. The Roman fast scan directions  use the default Teledyne configuration that alternates. In this figure, the fast scanners run from left to right in the odd numbered white colored outputs and from right to left in the even numbered grey colored outputs. The pixel readout rate is $2\times10^5~\textrm{pixels}~\textrm{s}^{-1}~\textrm{output}^{-1}$. There is a ~few pixel time overhead at the end of each row before clocking the first pixel of the next row. The prototype H4RG-10 shown in panel a is included for illustrative purposes only. It is not the flight design and some details have been deleted to comply with U.S. legal requirements. This figure should not be relied upon for engineering purposes.}
    \label{fig:h4rg_references}
\end{figure}

When seen in science data, the HxRG outputs usually appear as thick vertical stripes running the full ``height'' of the detector. For example, in an H4RG read out using 32 outputs, the pixels from one output would appear in a contiguous 128 pixel wide by 4096 pixel high rectangle. If one looks carefully at an exposure with some light in it, the reference pixels will appear as a four pixel wide dark frame surrounding the illuminated area (unless the pipeline has cropped them off). The description given here is how the outputs most often appear in science data. Sometimes they appear transposed, but this is a detail of how the instrument builders have chosen to present the data.

Fig~\ref{fig:h4rg_references} shows how these features are physically laid out looking down on a Nancy Grace Roman Space Telescope (``Roman'') H4RG. The important point here is that there are \textit{reference rows} and \textit{reference columns}.

The pixels in the reference rows are usually robustly averaged to apply an additive correction for constant offsets between outputs. Within SIRS, they are used in the same way as in many other pipelines. The reference rows correction is made at $f=0~\textrm{Hz}$, while the reference columns are used to correct all other frequencies. In Fourier space, the 0~Hz basis vector is orthogonal to the $f>0~\textrm{Hz}$ basis vectors, so both SIRS and reference rows corrections can be made without greatly affecting one another.

The reference columns and rows do different but complementary things. Only the left and right-most outputs in Fig~\ref{fig:h4rg_references} contain reference columns. Although in practice we find that there is often a useful degree of correlation with the other outputs. For every row that is read out, one gets eight reference samples from the reference columns. At Roman's 200~kHz pixel rate, the corresponding Nyquist frequency is about, $f_c\approx 741~\textrm{Hz}$. SIRS uses the reference columns to correct for frequencies $f>0~\textrm{Hz}$ and legacy reference rows techniques to correct for $f=0~\textrm{Hz}$.

In the rest of this paper, we focus on the reference columns correction. We refer the reader who is interested in legacy reference rows techniques to the literature. The online JWST user manuals\cite{JDox} describe what is arguably the state-of-the-art.
 
\subsection{SIRS Works with Common HxRG Clocking Patterns}\label{sec:clocking}

SIRS places no special requirements on how the detector is operated. It works with all of the readout schemes that are described in Teledyne's user manuals except for those that interleave guide windows (we are working on enhancements to accommodate guide windows now). For current Roman testing, we are using Teledyne's ``Enhanced Clocking Mode'' and ``Pixel by Pixel Reset''. The WFI's 32 H4RG outputs are read out using a Gen-III Leach Controller from Astronomical Research Cameras, Inc., at $2\times10^5~\textrm{pixels}~\textrm{s}^{-1}~\textrm{output}^{-1}$. There is a little time overhead at the end of each line to allow for clocking to and starting the next line. This overhead is currently about 7 pixel readout times.


Teledyne's HxRG detectors also offer global and line-by-line reset options. The Goddard Detector Characterization Laboratory's (DCL) experience has been that global reset does not work well for low background space astronomy. Early in JWST testing, and subsequently for Roman, the DCL reported that global reset was causing undesirable artifacts (more than one) \vs requirements. Because we had no compelling reason to use global reset, we quickly moved on to the other reset options.

The trade between pixel-by-pixel and line-by-line reset is more complicated. Both seem to work well for our applications. For JWST NIRSpec, we eventually selected pixel-by-pixel reset because: (1) each pixel sees the same lag between being reset and the first science sample, (2) each pixel sees the same exposure time independent of position, and (3) better thermal stability on account of the constant clocking cadence. At the time of writing, Roman is using pixel-by-pixel reset for the same reasons, although it is conceivable that this could change before flight.

In any case, we do not believe that the reset mode has any bearing on SIRS assuming that adequate settling time is allowed after resetting pixels. The key requirement is that the pixels always be sampled with the same constant and repeatable cadence apart from the end of line overhead. This requirement can be met with any of Teledyne's reset options.

In practice, there are many ways to implement Teledyne's recommended clocking patterns. In our software, the implementation-dependent parameters are defined in the module \texttt{SIRS.jl}. We recommend that users check with their instrument support teams to adjust these settings as necessary.

\section{Read Noise Characteristics}\label{sec:noise}

This section describes the read noise that SIRS was designed to remove. The test configuration differs from flight, and therefore what is said here will need to be revised when flight-like systems become available. In the following, we describe the read noise's spectral properties and covariance.

Our focus in this section is on the measured properties of the noise that SIRS was designed to correct. We refer readers who are interested in the underlying physical noise mechanisms to Rauscher (2015).\cite{Rauscher2015}

\subsection{Noise Power Spectrum}\label{sec:npsd}

\begin{figure}[htbp]
\centering
\includegraphics[width=\textwidth]{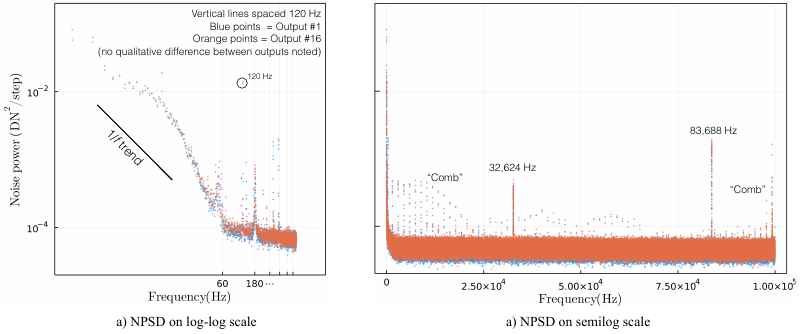}
\caption{Panel a) shows the power spectrum on a log-log plot, in instrumental units, for two different outputs. We saw no qualitative differences between any outputs. Instrumental units are sufficient for making relative comparisons. SIRS is highly effective in rejecting the $1/f$ noise that appears at low frequency. Other notable features include bands at 60~Hz and harmonics due to the U.S. power grid and b) a ``comb'' pattern spaced at intervals of the line frequency arising from interpolation over the regularly spaced end-of-line gaps. Panel b) shows the same points on a semilog scale. In addition to the features already mentioned, there is a line (of undetermined origin) at 83,688~Hz and its out of band second harmonic appears aliased down at 32,624~Hz. We looked for its third harmonic aliased down to 51,064~Hz, but did not find it. This suggests that either it is not very strong, or it may have been blocked by filters in the readout electronics. We tentatively attribute the enhanced ``comb''  pattern near 100~kHz to out of band ``comb'' aliased down.}
\label{fig:npsd}
\end{figure}

Figure~\ref{fig:npsd} shows the average noise power spectrum for two outputs of one flight-grade Roman H4RG tested using a DCL Leach Controller. The two outputs fall essentially on top of one another and we saw no major differences between any of the outputs. Section~\ref{sec:test} describes the test configuration in more detail.

To make these plots, we started with one 60 frame up-the-ramp sampled dark integration. We least squares fitted straight lines to all pixels and then subtracted the fits leaving residuals which we identified as read noise. For each frame; we reshaped the outputs as time ordered vectors, interpolated over gaps and statistical outliers, and computed the Fast Fourier Transform (FFT). The power spectrum is the absolute value of the FFT squared. The plots show the power spectrum averaged over all 60 frames.

\subsection{Covariance Matrix}\label{sec:V}

\begin{figure}[htbp]
\centering
\includegraphics[width=\textwidth]{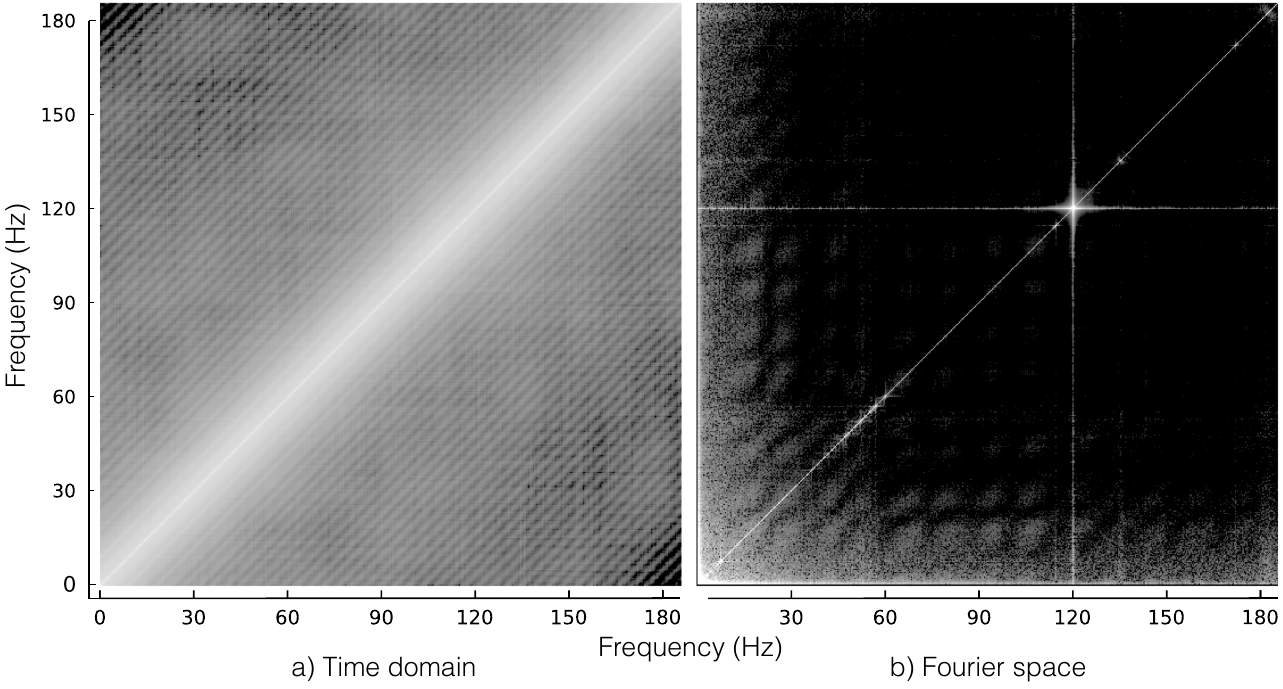}
\caption{Panel a) shows the covariance of the row-medians of output \#16 computed in the time domain. There is strong off-diagonal covariance, indicating that the different time steps (pixel indices) are correlated in this representation. The diagonal emerges much more clearly in b) Fourier space. Both panels a and b are on the same greyscale. The clouds around the edges in panel b reflect at least in part the uncertainties in the measurement. They become less pronounced as more data is added. The clouds are caused by the $1/f$ noise that SIRS aims to remove. The feature at 120~Hz requires more study, but it may appear so clearly because this line is so much stronger than any others (see Figure~\ref{fig:npsd}).}
\label{fig:covariance}
\end{figure}

Figure~\ref{fig:covariance} shows the low frequency portions of the empirical covariance matrices, $\mathbf{V}$,  for the medians of lines in output \#16. We computed $\mathbf{V}$ in both time and Fourier space using the same input data. We limited the frequency ranges for presentation purposes as it makes the diagonals easier to see. The other outputs are similar.

We computed covariance using the standard relation,
\begin{equation}
\mathbf{V} = \frac{\left(\mathbf{D}-\left<\mathbf{D}\right>\right)^{\rm H} \left(\mathbf{D}-\left<\mathbf{D}\right>\right)}{{n-1}},\label{eq:V}
\end{equation}
where angle brackets denote the expectation value, $n$ is the number of frames, and the symbol $^{\rm H}$ denotes the conjugate transpose. The factor $n-1$ appears in the denominator because we are estimating means from the data, $\mathbf{D}$.

To make $\mathbf{D}$, we started with a set of 100 identical 60 frame sampled up-the-ramp dark integrations. For each one, we fitted and subtracted straight lines to find the residuals which we identified as read noise. We then extracted output \#16 from every frame in the data set yielding 6,000 instances of the read noise in output \#16. Since we are most interested in low frequencies, we computed the median of each line as a statistically robust estimate of its expectation value. This yielded 6,000 vectors, each 4088 elements long (excluding the 4 reference rows at the top and bottom). To reject constant offsets between frames, we subtracted the mean of each vector from itself. For the Fourier space  matrix, we computed the FFT of each vector. In both cases, we put these vectors into a $6,000\times4088$ element matrix, $\mathbf{D}$, and computed $\mathbf{V}$ using Equation~\ref{eq:V}.

Figure~\ref{fig:covariance} reveals the key insight behind both SIRS and JWST's \IRSSquare.\cite{Rauscher2017}. The covariance matrix is more nearly diagonal in Fourier space. The clouds along the bottom and left edges of Figure~\ref{fig:covariance}b are due at least in part to uncertainties in the measurement. They become less pronounced when more data are used to make the measurement. Because the covariance matrix is roughly diagonal in Fourier space, we pay a much smaller penalty for ignoring the off diagonal covariance than we would if working in the time domain -as most pipelines do today.

\section{Theory}\label{sec:theory}


SIRS has the same mathematical foundation as \IRSSquare.\cite{Rauscher2017} The data are viewed as time series (also called ``streams'' here) rather than images. Building on the principal component analysis (PCA) that was done for JWST NIRSpec,\cite{Moseley2010a} the readout system is assumed to be covariance stationary during the frame readout time.

With these assumptions, Fourier space approximates the eigenspace and  thereby provides an uncorrelated representation of the read noise. The covariance matrix is nearly diagonal in Fourier space. One practical consequence is that we can fit the Fourier frequencies individually.  In Section~\ref{sec:test}, we show by test that these are reasonable assumptions for Roman H4RGs.

One might imagine that one could fit equally well in the time domain.  This would, however, require fitting full matrices rather than just vectors on account of the off-diagonal covariance. Doing so requires $\approx\frac{n}{2}\times$ more data, where $n$ is the length of the diagonal. For Roman's H4RGs, $n=4096$.  The stationary condition not only saves some computing time, but more importantly it allows us to work with far less data. Although computer time is inexpensive and becoming less so every year, observing time on space observatories will always be precious.

\subsection{Derivation of Key Equations}\label{sec:derivation}

This  derivation leads to almost the same equations as appear in Rauscher~\etal (2017)\cite{Rauscher2017}. However, we eliminate a filter that is no longer needed and introduce a  more compact notation that better reflects the references that are used here. Compared to our earlier work, this derivation has the advantage that it can be more easily generalized to handle more reference streams. It also lends itself to purely numerical implementation, which may be advantageous when many reference streams are used.

Each frame of data contains normal pixels (grouped by output), reference columns on the left and right sides, and reference rows on the top and bottom. The data from the normal pixels and reference columns can be formatted into time series vectors; (\bnm{n}) normal pixels, (\bnm{l}) left reference columns, and (\bnm{r}) right reference columns. Projected into Fourier space these become $\mathbcal{n}$, $\bm{\ell}$, and $\mathbcal{r}$. In this paper, boldface type indicates vectors and calligraphy fonts generally indicate Fourier space. Vectors that appear in calligraphy fonts are therefore complex.


There are gaps in the time series that complicate projection into Fourier space. The gaps arise because of the time overheads  (typically a few pixel readout times) required to start clocking new lines and because the normal pixels are sampled much more regularly than the reference columns. To accommodate the gaps, we use the incomplete real Fourier transform (IRFT) rather than the more common Fast Fourier Transform (FFT). Our IRFT is a specific type of nonuniform discrete Fourier transform\cite{Bagch1999} that avoids interpolation. We refer the interested reader to Appendix~\ref{sec:irft} for more information. It is not necessary to understand the IRFT's implementation to understand the  SIRS algorithm.

Once in Fourier space, we model the read noise in the normal pixels as a linear combination of the read noise in the references,
\begin{equation}
\bm{\alpha}\circ\bm{\ell} + \bm{\beta}\circ\mathbcal{r}=\mathbcal{n}.\label{eq:lin-mod}
\end{equation}
In this expression, $\bm{\alpha}$ and $\bm{\beta}$ are vectors of frequency dependent gains. The symbol $\circ$ represents the element-wise (Hadamard) product. SIRS uses least squares to find the optimal values of $\bm{\alpha}$ and $\bm{\beta}$ from a large set of training darks. A ``dark'' is an up-the-ramp sampled exposure with the detector blanked off so that it is not exposed to light. We typically use 100 up-the-ramp sampled darks, each having 60 frames, yielding 6,000 frames of data.

Because the Fourier components are uncorrelated, we can solve for them independently. Let $\msc{n}$, $\ell$, and $\msc{r}$ represent one element each of their corresponding vectors. Using Equation~\ref{eq:lin-mod}, we can write a set of $m\gg 2$ equations in the two unknowns, $\alpha$ and $\beta$.
\begin{align*}
\alpha\ell^1 + \beta\msc{r}^1 &= \msc{n}^1\numberthis \label{eq:m-eqns}\\
\alpha\ell^2 + \beta\msc{r}^2 &= \msc{n}^2\\
&\vdots\\
\alpha\ell^m + \beta \msc{r}^m &= \msc{n}^m
\end{align*}
In matrix notation, these become,
\begin{equation}\label{eq:m-matrix}
\begin{pmatrix}
\ell^1& \msc{r}^1\\
\ell^2& \msc{r}^2\\
\vdots& \vdots\\
\ell^m& \msc{r}^m
\end{pmatrix} 
\begin{pmatrix}
\alpha\\
\beta
\end{pmatrix} = 
\begin{pmatrix}
\msc{n}^1\\
\msc{n}^2\\
\vdots\\
\msc{n}^m
\end{pmatrix}.
\end{equation}
The least squares solution of Equation~\ref{eq:m-matrix} is,
\begin{equation}\label{eq:soln}
\begin{pmatrix}
\alpha\\
\beta
\end{pmatrix} = 
\begin{pmatrix}
\ell^1& \msc{r}^1\\
\ell^2& \msc{r}^2\\
\vdots& \vdots\\
\ell^m& \msc{r}^m
\end{pmatrix}^+
\begin{pmatrix}
\msc{n}^1\\
\msc{n}^2\\
\vdots\\
\msc{n}^m
\end{pmatrix},
\end{equation}
where the $^+$ sign denotes the Moore-Penrose inverse of the ``$\ell \msc{r}$ matrix''. Because the $\ell \msc{r}$ matrix's columns are linearly independent (they are physically different references), the Moore-Penrose inverse has an analytic solution. Substituting it, Equation~\ref{eq:soln} simplifies to,
\begin{equation}\label{eq:soln2}
\begin{pmatrix}\alpha\\ \beta\end{pmatrix} =
\begin{pmatrix}
	\ell_i^*\ell^i& \ell_i^*\msc{r}^i\\
	\msc{r}_i^*\ell^i& \msc{r}_i^*\msc{r}^i
\end{pmatrix}^{-1}
\begin{pmatrix}\ell_i^*\msc{n}^i\\ \msc{r}_i^*\msc{n}^i\end{pmatrix}.
\end{equation}
In these expressions, the $^*$ symbol denotes complex conjugation. We have also used the Einstein summation convention to achieve a clean notation. For example, $\ell_i^*\ell^i$ stands for $\sum_{i=1}^m\ell_i^*\ell^i$.

To further simplify Equation~\ref{eq:soln2}, it is helpful to define some sums;
\begin{align}
\mbb{L} &= \ell_i^*\ell^i,\label{eq:L}\\
\mbb{R} &= \msc{r}_i^*\msc{r}^i,\\
\mbb{X} &= \msc{r}_i^*\msc{n}^i,\\
\mbb{Y} &= \ell_i^*\msc{n}^i, \textrm{and}\\
\mbb{Z} &= \ell_i^*\msc{r}^i.\label{eq:Z}
\end{align}
The first two sums are real and the last three are complex. With these substitutions, Equation~\ref{eq:soln2} simplifies to,
\begin{align}
\alpha &= \frac{\mbb{R}\mbb{Y}-\mbb{X}\mbb{Z}}{\mbb{R}\mbb{L}-\mbb{Z}\mbb{Z}^*}\textrm{~and}\label{eq:alpha}\\
\beta &= \frac{\mbb{X}\mbb{L}-\mbb{Y}\mbb{Z}^*}{\mbb{R}\mbb{L}-\mbb{Z}\mbb{Z}^*}\label{eq:beta}.
\end{align}

In Equation~\ref{eq:lin-mod}, we defined $\bnm{\alpha}$ and $\bnm{\beta}$ as vectors in Fourier space, but Equations~\ref{eq:alpha}-\ref{eq:beta} are for single Fourier components. Equations \ref{eq:L}-\ref{eq:beta} must therefore be evaluated for all frequencies that will be used to make the reference correction.

There are other ways to find the least squares solution to Equation~\ref{eq:m-matrix}. The least squares itself is just a generalization of the familiar solution for fitting a 2-parameter straight line to data. We could therefore have subtracted the right side of Equations~\ref{eq:m-matrix} from the left side, summed the squares of the deviates (remembering that the deviates are complex numbers), and minimized this sum over $\alpha$ and $\beta$. The result is the same, although the geometric approach presented here scales more gracefully as additional reference streams become available.

\subsection{Example $\alpha$ and $\beta$}\label{sec:examining-alpha-beta}


We developed SIRS using data from 18 Roman flight candidate H4RGs. Some of these tests are described in more detail in Section~\ref{sec:test}. Here we briefly discuss the measured values of $\bnm{\alpha}$ and $\bnm{\beta}$ for one Roman H4RG flight candidate: SCA 20663. These results are broadly representative.

Although $\bnm{\alpha}$ and $\bnm{\beta}$ clearly differ from detector to detector, the general characteristic of $1/f$ noise with spectral lines at 60~Hz and harmonics is always seen. In practice, SIRS must be trained for every detector. For any one detector, each test lasted about 8~hours. We detected no time dependent changes in $\bnm{\alpha}$ and $\bnm{\beta}$ for any of these detectors.

\begin{figure}[h]
\centering
\includegraphics[width=\textwidth]{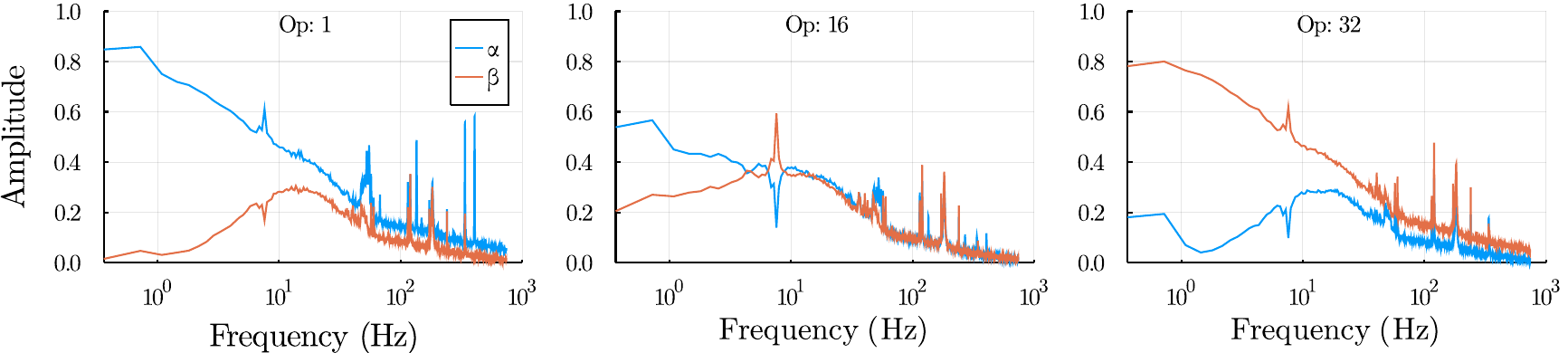}
\caption{This figure shows $\bnm{\alpha}$ and $\bnm{\beta}$ for 3 of H4RG 20663's 32 outputs. Output \#1 is the left-most output, output \#16 is just left of center, and output \#32 is the right-most output. As expected, the relative weights taken by $\bnm{\alpha}$ and $\bnm{\beta}$ shift depending on which reference columns are closest. Broadly speaking, outputs 2-15 look intermediate between outputs 1 \& 16 and similarly outputs 18-31 look intermediate between outputs 17 \& 32.}
\label{fig:sirs-alpha-beta}
\end{figure}

Fig~\ref{fig:sirs-alpha-beta} shows the low frequency behavior on a semi-log scale as a function of output number ($\bnm{\alpha}$ and $\bnm{\beta}$ are calculated separately for each output). The H4RG's outputs are numbered 1--32, with output \#1 being closest to the left reference columns that $\bnm{\alpha}$ operates on and output \#32 being closest to the right reference columns that $\bnm{\beta}$ operates on. Output \#16 is just to the left of the array's center line.


The HxRG's reference pixels are physically located in the ROIC where they appear to be in images. As one would expect from the physical proximity, the relative importance of $\bnm{\alpha}$ and $\bnm{\beta}$ shifts depending on which reference columns are closest. Although the noise is dominated by $1/f$ at low frequencies, many spectral bands and lines are present. We have not tried to track down all of their causes, but we find that cryocoolers often appear at about 8~Hz, as seems to be the case here.

\begin{figure}[h]
\centering
\includegraphics[width=\textwidth]{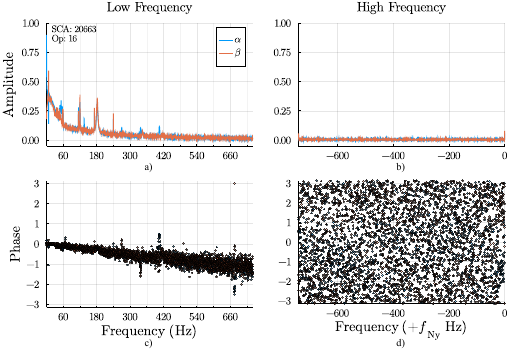}
\caption{a) This figure re-plots Fig~\ref{fig:sirs-alpha-beta} (center) on a linear scale. Visible features include $1/f$ noise and lines and bands associated with the US power grid at 60, 120, 180, and 240~Hz. c) Because the reference pixels are the first pixels read in every line, the absolute value of the phase increases from zero at 0~Hz to $\frac{\pi}{2}$ at $f_\textrm{Ny,line}$. b) There is very little amplitude near the Nyquist frequency, $f_\textrm{Ny}=200~\textrm{kHz}$, and d) phase is almost entirely noise at these frequencies. We therefore zero-out frequencies $f>f_\textrm{Ny,line}$ when working with the existing Roman data.}
\label{fig:sirs-wplot}
\end{figure}

Fig~\ref{fig:sirs-wplot} shows the same data on a linear scale and includes the highest frequencies near $f_\textrm{Ny}=200~\textrm{kHz}$. Although we only show one output, these plots are typical if allowance is made for the shifting relative importance of $\bnm{\alpha}$ and $\bnm{\beta}$.

Fig~\ref{fig:sirs-wplot}a again shows the strong low frequency correlation. It also highlights some features that one often sees with US power: lines at 60~Hz (and harmonics) with a strong band at 180~Hz on account of the three phase power and fluorescent lights in the lab.  Panel c shows that at 0~Hz, everything is in phase. But, because the reference pixels are the first pixels read on every line, the absolute value of the phase increases to $\frac{\pi}{2}$ at $f_\textrm{Ny,line}$. The phase noise increases as the amplitude decreases.

Panels b and d show the high frequency behavior near $f_\textrm{Ny}=200~\textrm{kHz}$. Our conclusion is that there is no useful reference information at these frequencies for these detectors. We are therefore zeroing-out $\bnm{\alpha}$ and $\bnm{\beta}$ for  $f>f_\textrm{Ny,line}$ in the current Roman test data.

This was a pleasant surprise. In JWST NIRSpec, there was significant ACN at $f_\textrm{Ny}$. This mixed with low frequency $1/f$ to produce a high frequency peak (See Fig~8a of Ref.~\citenum{Rauscher2017}). The ACN was caused by a Teledyne-proprietary design choice in the H2RG column buses.\cite{Rauscher2015,Rauscher2017} Roman's H4RG-10s do not seem to have this spectral feature. Although we do not have the same insight into the relevant details of the H4RG's ROIC design, Rauscher did discuss ACN with Teledyne while they were designing the H4RG and what could be done to improve it.

Because we thought that similar ACN might still exist in the Roman H4RGs, SIRS fits for both low and high frequencies as shown here. If we had known in advance that only the low frequencies were important, it would have simplified the implementation somewhat. For now, we have left this feature in as things may change when the detectors are mated to the real flight electronics.

In any event, we do not expect SIRS to be particularly powerful at suppressing ACN. Unlike JWST's \IRSSquare, which uses a specialized clocking pattern to fully sample ACN, Conventional clocking patterns do not sample reference often enough to be very effective at suppressing ACN.

\section{SIRS in Practice}\label{sec:computers}

Two steps are required to use SIRS. First, SIRS must be trained using a large set of up-the-ramp sampled darks (see Section~\ref{sec:test}). We call this the ``frontend''. It is computationally intensive, but the SIRS package provides a function for saving the trained state. Once trained, using SIRS to calibrate new data (the ``backend'') is about as computationally demanding as conventional reference correction. We typically work on at least a small server, but believe that it would be practical to run the backend on even a scientific laptop.

\begin{figure}[htbp]
\centering
\includegraphics[width=\textwidth]{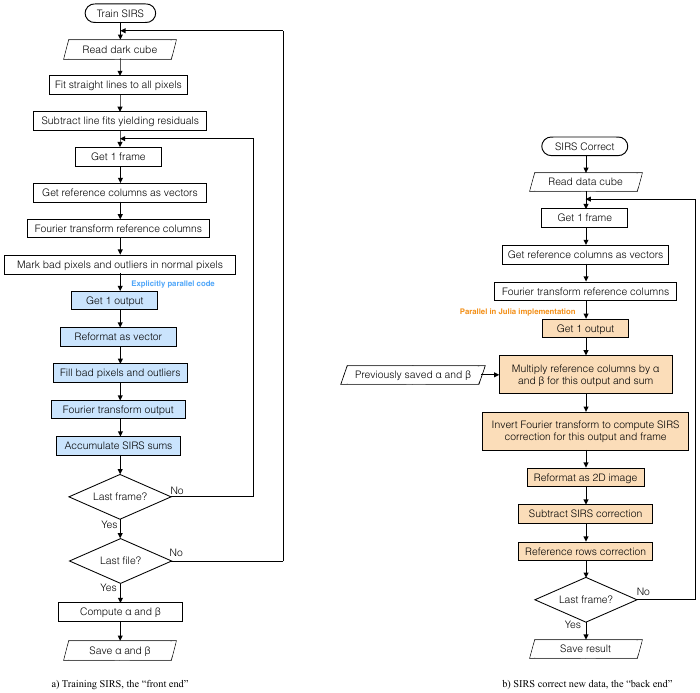}
\caption{To SIRS correct data, one must first a) train the algorithm using a large set of training darks and then b) apply the SIRS correction to the new data. We refer to part a) as the ``frontend'' and b) as the ``backend''. Only the frontend is computationally intensive. It includes a box labeled, ``Accumulate SIRS sums''. This evaluates Equations~\ref{eq:L}-\ref{eq:Z}. The ``Compute $\alpha$ and $\beta$'' box solves Equations~\ref{eq:alpha} and \ref{eq:beta}. In the Julia language implementation, parts of both the front and backend use thread based parallelism. This is indicated by shading in some boxes. The python language backend is single threaded, although in practice some steps are nevertheless parallelized by python's numerical packages.}
\label{fig:flowchart}
\end{figure}

Fig.~\ref{fig:flowchart} provides a flowchart showing how the equations of Section~\ref{sec:theory} are used. To better understand the Julia implementation details, we recommend examining the source code, and especially the \texttt{coadd.jl} module. For those using Jupyter notebooks, the notation is the same as that used here. The SIRS distribution also contains a PDF of \texttt{coadd.jl} (in the Documentation directory) to aid those not using Jupyter.

We have successfully used the frontend for calibrating JWST H2RG detectors ($\rm 2K\times2K$~pixels) on a small Linux server. The server has 8$\times$3.5~GHz cores and 256~GB of RAM. The execution time for training SIRS for one H2RG was about an hour. Based on our experience, we believe that 8$\times$3.5~GHz cores and 64~GB of RAM is probably about the minimum desirable configuration for H2RG data.
 
For Roman's H4RGs ($\rm 4K\times4K$~pixels), we developed SIRS using one node of the National Center for Climate Simulation (NCCS) high performance Prism Cluster.  Each Prism node has 40-cores capable of running at up to 3.9~GHz (max). Although Prism provides up to 4$\times$ NVIDIA V100 GPUs per node, we did not use the GPUs because of a desire to use 64-bit floating point arithmetic (the GPUs support up to 32-bits natively).

We used  NCCS's ``Ganglia'' utility to monitor Prism while training. Typical execution time was about 90~minutes. The peak RAM utilization was a little over 128 GB. The CPUs were typically running at about 800\% utilization, although the load sometimes dropped as low as 100\% during I/O operations. It peaked occasionally at 4000\% for tensor operations that were able to full utilized the Intel Math Kernel Library's (MKL) optimization. In practice, a good configuration for training SIRS for H4RG data would have $\geq$8 fast cores and 256~GB of RAM.

For both JWST H2RG and Roman H4RGs, we have found SIRS to be computationally stable. We have run the frontend on 15 different Roman H4RGs so far. To within the uncertainties, the computed values of $\bm{\alpha}$ and $\bm{\beta}$ are always similar to those shown in Figures~\ref{fig:sirs-alpha-beta} and \ref{fig:sirs-wplot}. When allowance is made for the higher noise, this is also true when different subsets of training data are used for each detector.

\section{Validation using Roman Test Data}\label{sec:test}

To validate the reduction of correlated noise using the SIRS technique described in Section \ref{sec:theory} and verify no changes in measured response, we compare the noise properties of a set of SIRS corrected data and a set of data corrected with a standard choice of reference pixel correction. For the standard choice of reference pixel correction, we choose to correct each output by the median of the last 4 rows of reference pixels. This is a common choice of reference pixel correction that is used in the analysis of the Roman mission's sensor acceptance testing.

The set of data used in these validation tests are from the Roman flight candidate detector 20663 taken by Goddard's Detector Characterization Lab (DCL) as part of the sensor's acceptance test. All data was taken using Astronomical Research Camera Inc.'s Gen-III Leach controllers with a ARC-22 timing card and ARC-46 eight channel video processing boards. The H4RG is read out using line-by-line reset clocking patterns for 32 output mode described in the H4RG-10 user manual implemented with custom DCL software. The H4RG is cooled to 95~K with a cryocooler and operated at a 1~V reverse bias for these tests. The DCL also tested this detector at 90~K and 0.5~V reverse bias. SIRS worked comparably well. The results shown here are representative.

The data for the noise comparison are a set of 105 dark exposures, consisting of 60 consecutive reads each. This set of data is typically used to estimate the total slope noise of a detector. Dark current is known to be very low ($\lesssim$0.001~e-/s/pixel) for these devices from qualification testing and acceptance tests results. The second data set to look for any changes in measured response is a set of 20 exposures consisting of 100 consecutive reads with 1.4~$\mu$m illumination using a tungsten lamp directed through a monochromator to produce a flatfield onto the detector.

All the following analysis also excludes pixels ($\sim$1.2\%) that have been flagged as inoperable according to the results of Goddard Detector Characterization Lab Acceptance Testing.

\subsection{Noise comparison}

\begin{figure}[h]
\centering
\includegraphics[width=\textwidth]{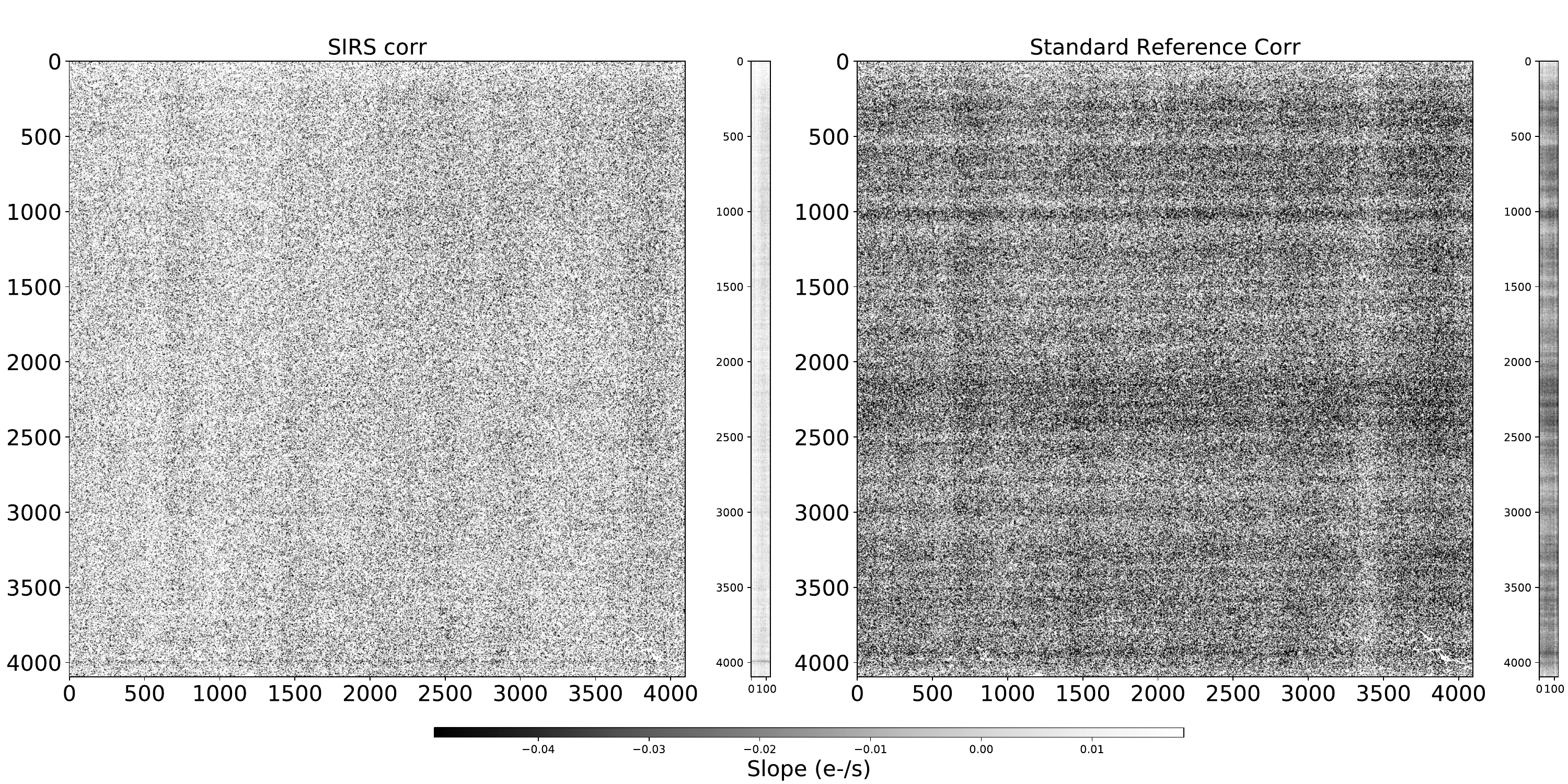}
\caption{Representative comparison of slope images derived from (left) SIRS corrected ramp and a standard reference subtraction corrected ramp (right). Beside each image the output median for the 32 outputs is plotted to highlight the horizontal banding. The slope image derived from the SIRS corrected data show substantial reduction in the horizontal banding. The greyscale in the images is up-the-ramp group slope in units $e^-~\textrm{s}^{-1}~\textrm{pixel}^{-1}$. Since submitting this article, a JWST colleague has pointed out that with careful tuning, one can approach SIRS' low frequency noise suppression using a reference columns option that is built into the JWST pipeline. We agree, but believe that that SIRS still offers less visible statistical advantages without any need for tuning.}
\label{fig:img-comp}
\end{figure}

We can qualitatively compare the amount of correlated noise in the SIRS corrected set of data and the standard corrected set of data by inspecting the slope images from a representative exposure. Fig~\ref{fig:img-comp} shows the slope image from fitting the 60 samples up-the-ramp of a single exposure for SIRS corrected up-the-ramp data and up-the-ramp data using the standard correction. In Fig~\ref{fig:img-comp}, the horizontal banding from correlated noise in the slow readout direction is evident and stronger in the data corrected using only the last 4 rows of reference pixels.

Both the SIRS and standard reference corrections were done frame-by-frame, prior to fitting for slope. Although reference correction can be done before or after slope fitting, and will often give about the same result, the two are not mathematically equivalent. We view reference subtraction as part of differentially sampling the detector and therefore do it before slope fitting.

For both SIRS and JWST's \IRSSquare, we are often asked if it is acceptable for the up-the-ramp samples to be averaged into ``groups'' before reference correction. To a first approximation, the answer is yes. Since the operations are linear (apart from outlier rejection), the result will be about the same to within the roundoff errors.

\clearpage
\begin{sidewaysfigure*}[!ht]
\centering
\includegraphics[width=\textwidth]{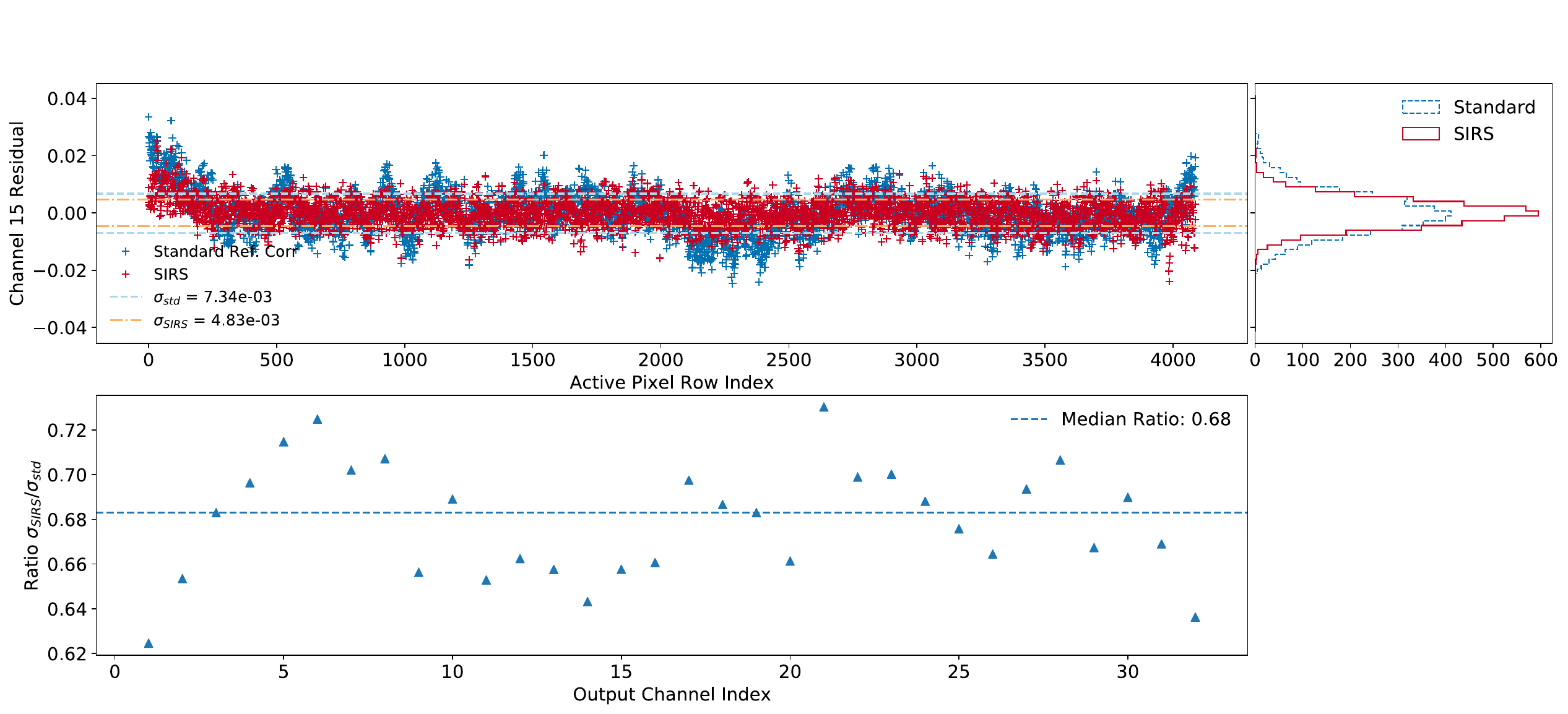}
\caption{Representative comparison of residuals in the median slope in output \#15 as a function of pixel row. Top panel shows the residual between the output median and the average value of the output median over all active pixel rows for both SIRS and standard reference pixel corrected data. Bottom panel shows the ratio of the standard deviation of the residuals in the top panel (SIRS divided by the standard correction) for all output channels. The residuals of the SIRS corrected slope image data are more strongly peaked at zero and show less variation than the data using the standard correction.}
\label{fig:resid-comp}
\end{sidewaysfigure*}
\clearpage

To quantify the excess noise in data using only the standard correction, we can examine the residual variation in each output along the slow readout direction after subtracting off the output mean slope. Fig~\ref{fig:resid-comp} shows the residuals of a central output (\#15) for both sets of data for a representative exposure in the top panel. On the lower panel of Fig~\ref{fig:resid-comp}, the ratio of the standard deviation of the SIRS corrected data residuals to data residuals using the standard correction is plotted for each output of the same exposure. For the exposure shown in Fig~\ref{fig:resid-comp}, the SIRS corrected slope image data has approximately 33\% less variation. Overall, when looking at a total of 105 exposures, the residual variation in the SIRS corrected data is about 25\% less than the residual variation in data corrected using the standard reference pixel correction.

\begin{figure}[h]
\centering
\includegraphics[width=\textwidth]{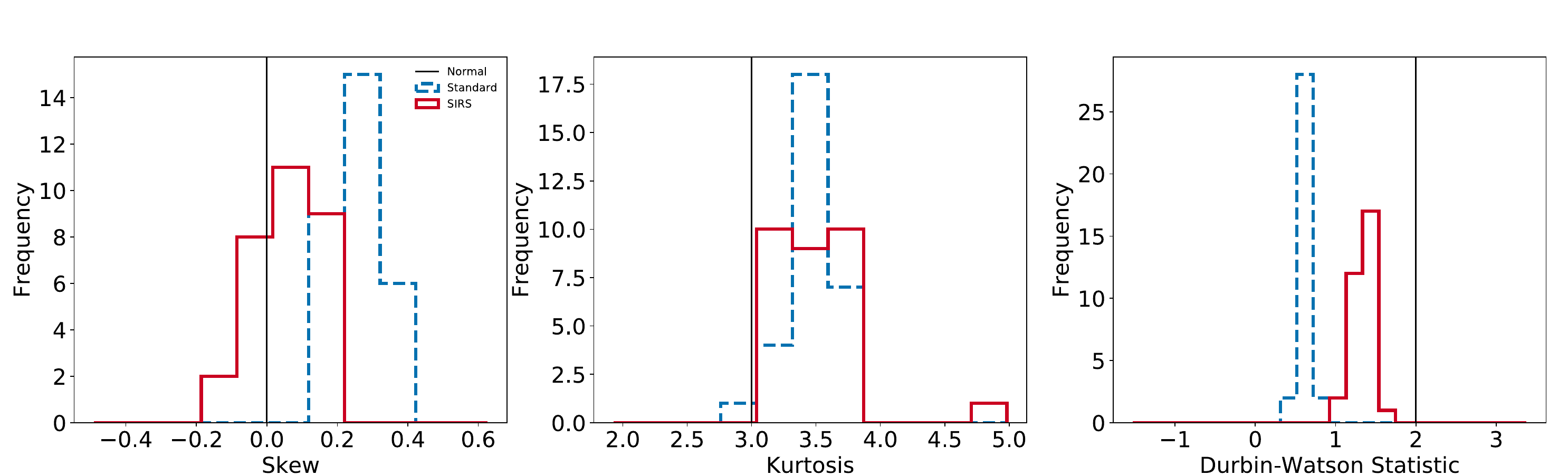}
\caption{Statistical properties for output residuals from Fig~\ref{fig:resid-comp}. Left: distribution of skewness calculated for each output's residuals. Middle: distribution of kurtosis calculated for each output's residuals. Right: distribution of Durbin-Watson statistic calculated for each output's residuals. For a normal distribution, skewness would be zero, kurtosis would be 3, and the Durbin-Watson statistic would be 2. The SIRS corrected data residuals have less skewness, data residuals from both corrections have excess kurtosis, and the SIRS corrected data have less auto-correlation.}
\label{fig:norm-comp}
\end{figure}

The histograms of the residuals per output channel in Fig~\ref{fig:resid-comp} appear coarsely normal, but to assess this quantitatively, we can look at statistical properties for both sets of residuals to test for normality. Normally distributed residuals would indicate random variation with pixel row and that the optimal model for the slope image data in an output is indeed a constant value independent of pixel row. Non-normality would suggest additional parameters are needed to describe the data. Fig~\ref{fig:norm-comp} shows the calculated skewness, kurtosis, and the Durbin-Watson statistic for the residuals in all 32 outputs as 3 histograms. Skewness and kurtosis are the commonly defined standardized third and fourth moments of a probability distribution, respectively. The Durbin-Watson statistic is a measure of the serial auto-correlation in the residuals\cite{Durbin-1950}. For a normal distribution, skewness would be zero, kurtosis would be 3, and the Durbin-Watson statistic would be 2 assuming no auto-correlation. For the representative exposure in Fig~\ref{fig:norm-comp}, the SIRS corrected data residuals have less skewness, data residuals from both corrections have excess kurtosis, and the SIRS corrected data have less auto-correlation. These results in conjunction with the results of Fig~\ref{fig:resid-comp} suggest that the residuals from both types of data are not completely normally distributed. But they suggest that the residuals from the SIRS corrected data could be represented by a Gaussian distribution with narrower standard deviation with enhanced tails.

\clearpage
\begin{sidewaysfigure}[h]
\centering
\includegraphics[width=\textheight]{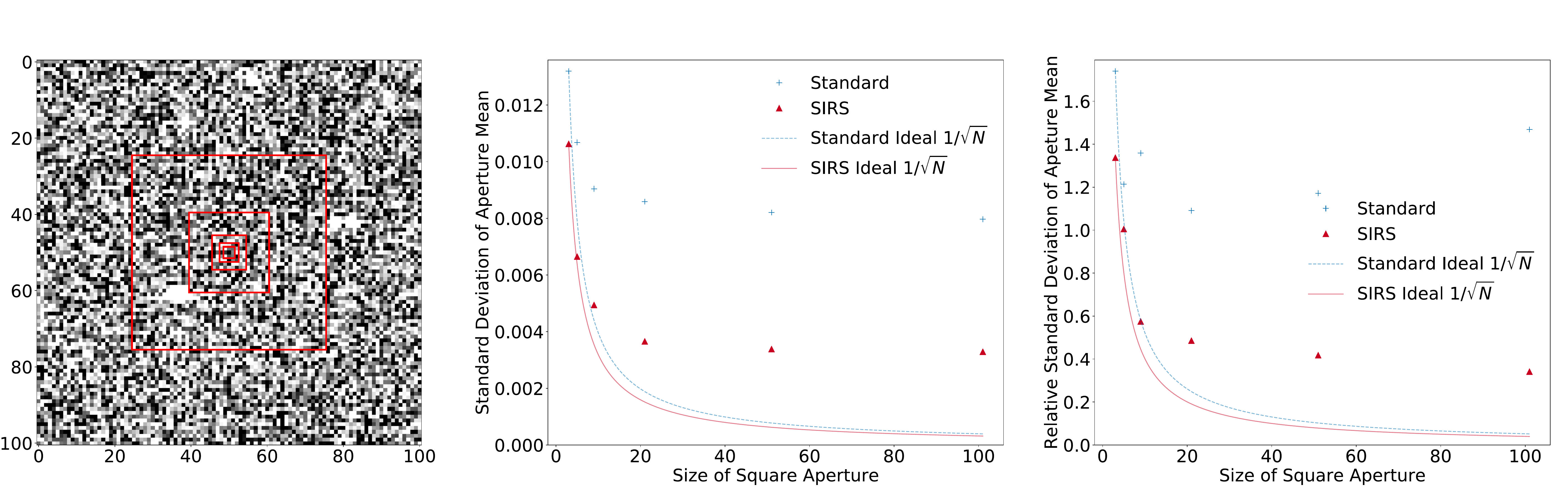}
\caption{Left: Image of the region (101 $\times$ 101) where the aperture averaging is done with red rectangles overlaying the smaller apertures. Middle: Standard deviation of the mean as a function of aperture, calculated using 105 exposures. Right: Relative standard deviation (standard deviation divided by absolute value of mean) as a function of aperture, calculated using 105 exposures. The variation (noise) of the aperture mean in the SIRS corrected slope data decreases more rapidly with aperture size and to nearly half the value of the variation of the aperture mean in the slope data. However, both corrections bottom out after the aperture is larger than 9~$\times$~9 indicating there is still room for improvement; this is likely due to residual $1/f$ noise that is not removed by the current algorithm. Nevertheless, the relative standard deviation is smaller for the SIRS corrected data, and it decreases below the mean value in contrast to the relative standard deviation for the data corrected using the standard correction.}
\label{fig:apnoise-comp}
\end{sidewaysfigure}
\clearpage

Normality is a key feature to look for in this data because it implies the pixel noise properties could permit averaging over multiple pixels to increase precision and accuracy by the reduction of uncorrelated noise. We compare how the pixel noise averages down for both types of reference pixel correction using square apertures of increasing size in the central output (\#15) of their respective slope images. The square aperture sizes are 3$\times$3, 5$\times$5, 9$\times$9, 21$\times$21, 51$\times$51, and 101$\times$101 pixels. The mean signal in an aperture is calculated for each exposure. The standard deviation of the mean as a function of aperture size is then calculated. The variation in the mean should decrease with increasing aperture size for data containing few correlations. Fig~\ref{fig:apnoise-comp} shows the results for computing aperture means for 105 exposures and estimating the standard deviation of that mean for each aperture across those exposures. The middle panel of Fig~\ref{fig:apnoise-comp} shows that the variation (noise) of the aperture mean in the SIRS corrected slope data decreases more rapidly with aperture size and to nearly half the value of the variation of the aperture mean in the slope data corrected with the standard correction. The SIRS corrected data, however, does still plateau at noise levels above the theoretical limit. This is likely due to residual $1/f$ noise not completely removed in this simple algorithm. The right panel of Fig~\ref{fig:apnoise-comp} shows the relative standard deviation (standard deviation divided by absolute value of mean). We can see that the relative standard deviation is smaller for the SIRS corrected data, and that it decreases below the mean value in contrast to the relative standard deviation for the data corrected using the standard correction. This is consistent with the larger amount of correlation in the data using the standard correction.

\subsection{Comparison to Subtracting Reference Columns}\label{sec:refcols}

Since submitting this paper, a JWST colleague has compared SIRS to one of the JWST pipeline reference correction methods using columns.\cite{JDox} Using the side reference columns and smoothing them over 11 rows, he was able to produced images that were visibly closer to Figure~\ref{fig:img-comp} (left) with regard to horizontal banding. Cosmetically, the result was much cleaner than just using the reference rows, and the visible difference was much less dramatic compared to Figure~\ref{fig:img-comp}.

Mathematically, this makes sense. Smoothing the columns with a kernel can be implemented with a filter in Fourier space. SIRS can also be viewed as applying (more complex) filters in Fourier space. Viewed this way, SIRS still enjoys compelling statistical advantages, even if the difference is less obvious to the eye.

The first is that SIRS is statistically optimal using least squares as the figure of merit. To the extent that the the deviates follow an approximately Normal distribution, it is also optimal with regard to maximum likelihood and bayesian figures of merit. The second is that SIRS does the tuning with superhuman precision. It is impossible to reproduce the complex structure shown in Figure~\ref{fig:sirs-wplot} using only a small handful of tuning parameters such as window height and gain. Finally, SIRS uses frequency dependent gain and phase information that is discarded in the implementations that we are aware of.

\subsection{Response comparison}
\begin{figure}[h]
\centering
\includegraphics[width=\textwidth]{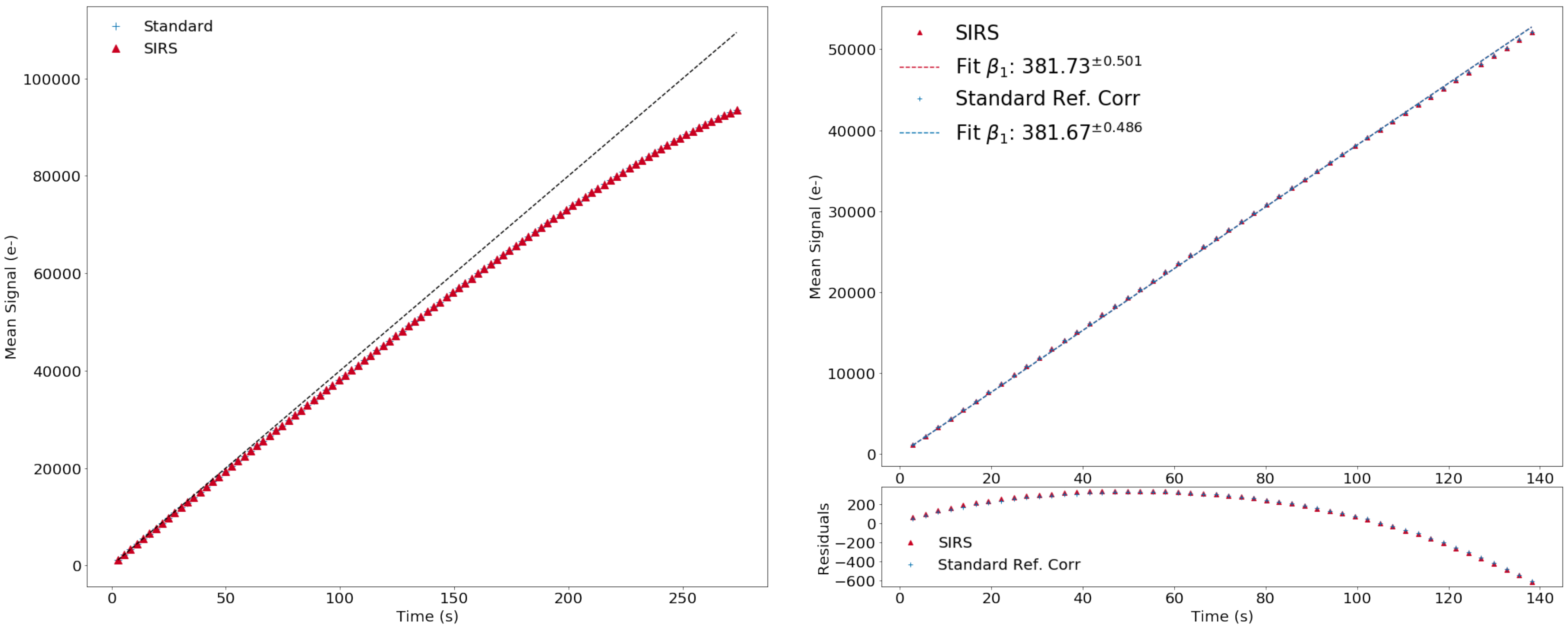}
\caption{Left: Mean signal in illuminated frames up-the-ramp for a representative illuminated exposure. The array mean is estimated in each frame after subtracting the baseline signal in the first frame and plotted as a function of time. The means for the SIRS corrected data lie on top of the means for the data corrected using the standard reference subtraction.The dashed black line shows the linear trend expected for constant signal. Right: The fits of the slopes and corresponding residuals for the mean signals of the SIRS corrected data and the data corrected using the standard reference pixel correction. The bottom panel shows the residuals to the fits. The response rolls off as the detector system nears saturation around 80~ke-, but there are no significant differences in response due to the SIRS correction. The best fit slopes, $\beta_{1,SIRS}: 381.73 \pm 0.501$ and $\beta_{1,std}: 381.67 \pm 0.486$, for both data sets are similar and indistinguishable within the error bars of the fit.}
\label{fig:lin-comp}
\end{figure}

To look for any potential differences in the measured response when using the SIRS correction versus the standard correction, we can examine the mean signal as a function of frames up-the-ramp using the second data set with flatfield illumination. The array mean is estimated in each frame after subtracting the baseline signal in the first frame and plotted as a function of time in Fig~\ref{fig:lin-comp}. No significant differences are apparent in the response implied from the two distinct reference pixel correction schemes.

\section{Future Work}\label{sec:future}

Going forward, we plan significant changes for Roman on account of the different clocking pattern that will be used for flight and additional references that will become available. For JWST, SIRS already appears to do a superior job suppressing $1/f$ banding compared to what is in the pipeline for NIRCam and NIRISS. However, the pipeline reference rows correction still appears to be somewhat better at suppressing alternating column noise (ACN). Future work for JWST will focus on reducing ACN.

\subsection{Planned Roman Upgrades}\label{sec:roman}

The DCL acquired the data that underlie this paper using flight candidate H4RGs and non-flight Leach controllers. For flight, most of the same detectors will be built into ``triplets''. A triplet consists of: (1) H4RG detector, (2) interconnect cable, and (3) ACADIA application specific integrated circuit (ASIC). The differences include the ACADIA replacing the Leach controller, a different clocking pattern, and additional references will be available.

The ACADIA is a new IR array controller ASIC that aims to improve on the Teledyne SIDECAR\cite{Loose2007} ASICs that were used by JWST, Euclid, and many other observatories. Because it is new, we cannot say anything about the performance of flight ACADIAs at this time other than we understand that they are meeting their requirements. For more information about ACADIA, we refer the interested reader to Loose~\etal (2018)\cite{Loose2018}.

Roman plans to use the H4RG's ``guide window'' for observatory guiding. Although the Roman Project has tested guide windows extensively, they were not used for the data that are the basis of this paper. Using the guide window requires a more complex clocking pattern that interleaves controlling the guide window with reading out normal pixels. The new timing pattern will have different gaps and overheads than the one that was used for this paper. This will in turn require modifications to the SIRS software. Depending on how extensive the modification are, SIRS may be superseded by something that uses the same mathematical techniques, but that is adapted to use all of the references available at triplet level optimally including the H4RG's reference output.

\subsection{Planned JWST Upgrades}\label{sec:jwst}

With JWST launch now imminent, we plan further work optimizing SIRS for JWST NIRCam and NIRISS. NIRSpec already provides \IRSSquare readout which will be superior to SIRS.

The primary limitation that we are aware of concerns the ACN that we have only briefly touched on here paper because it falls outside the scope of the Fourier domain reference corrections that are our main focus. At the time of writing, SIRS implemented a very simple version of reference rows correction to remove constant drifts between outputs. We believe that with a little more work on the reference rows correction, we should be able to produce a version of SIRS that is superior to JWST's pipelined reference correction in all respects.

\section{SIRS as a Simple Example of Supervised Machine Learning}\label{sec:ml}

SIRS has several characteristics that are typical of supervised machine learning. There is no physics and we never explicitly program SIRS to make optimal reference corrections. The quality of SIRS' correction depends entirely on the training data. If the characteristics of the detector system should change, then so will the training data, and so too will the SIRS correction that is applied.

Although SIRS assumes a linear mapping between the references and the normal pixels, everything about the mapping is inferred from the training data by SIRS itself using least squares to perform the optimization. As an optimizer, least squares has many advantages. Among these, it is linear and deterministic, giving a result that does not depend on the initial conditions of the optimization.

\section{Summary}\label{sec:summary}

We have described Simple Improved Reference Subtraction (SIRS). SIRS uses training data  with linear algebra and least squares to reference correct Teledyne HxRG detector systems. SIRS is statistically optimal with least squares as the figure of merit. For systems showing significant banding aligned with the fast scan directions, SIRS may reduce correlated noise including $1/f$ banding and some types of electromagnetic interference.

SIRS works with some of the most common HxRG readout patterns and makes no special demands on the hardware or how the detector is read out. When applicable, SIRS can be used to reduce the correlated noise of even archival data.

\appendix

\section{Incomplete Real Fourier Transform}\label{sec:irft}

The reference column time series have fewer elements than the normal pixels time series. This, and the time overhead for starting new lines, creates gaps in the time series. We account for these gaps using incomplete real Fourier transforms (IRFT). We define IRFT to be the orthogonal projection of a real valued data vector, $\mathbf{d}$, into a lower dimensionality Fourier space using only the available samples.

SIRS uses Julia's FFTW package (https://juliahub.com/docs/FFTW) for computing Fourier transforms. In turn, Julia's FFTW package provides bindings to the C language FFTW library for fast Fourier transforms (http://www.fftw.org). Our implementation of IRFT uses the Moore-Penrose inverse of a matrix containing the FFTW Fourier basis vectors to find the least squares solution. Specifically,
\begin{equation}
\textrm{IRFT}\left(\mathbf{d}\right)=\mathbf{B}^+ \mathbf{d},\label{eq:irft}
\end{equation}
where $\mathbf{B}$ is the incompletely sampled Fourier basis matrix.

Each column of $\mathbf{B}$ is a Fourier basis vector, evaluated only at the rows for which $\mathbf{d}$  contains data. By inspection of FFTW's online documentation (the C language version), FFTW's basis vectors are,
\begin{equation}
B^j_k = e^{2\pi(j-1)(k-1)\sqrt{-1}/n}.\label{eq:B}
\end{equation}
This expression includes a correction for Julia using unity offset arrays whereas C uses zero-offset arrays. In Eq.~\ref{eq:B}, $j$ is the row index (equal to the time step in the clocking pattern for one output), $k$ is the column index, and $n$ is the total number of time steps including all gaps in one output of one frame of data.

In our earlier \IRSSquare paper,\cite{Rauscher2017} we interpolated over the gaps and computed FFTs. This is what was practical at the time given the available computers. Since then, computers have become more powerful. When computationally practical, we believe the IRFTs that are used here are superior. They eliminate the uncertainties (and sometimes extrapolations) that are needed when interpolating and computing FFTs.

\subsection*{Disclosures}
We have no relevant financial interests in the manuscript and no other potential conflicts of interest to disclose.

\subsection* {Acknowledgments}
This work was supported by NASA as part of the Nancy Grace Roman Space Telescope and James Webb Space Telescope Projects. Resources supporting this work were provided by the NASA High-End Computing (HEC) Program through the NASA Center for Climate Simulation (NCCS) at Goddard Space Flight Center. We are grateful for the invaluable support of the NASA Goddard Space Flight Center Detector Characterization Laboratory (DCL) team. The DCL produced all of the data used for this study. We wish to thank our JWST colleague, Dr. Chris Willott, of Herzbert Astrophysics, for comparing SIRS to the JWST pipeline. We wish to thank the referees for many helpful suggestions that have improved this paper. In particular, it was one of the referees who first mentioned intensity mapping (see Section~\ref{sec:sci}) as a potential application.

\vspace{2ex}
\noindent{\it Software: Julia,\cite{bezanson2017julia}  Python3\cite{python3}, Matplotlib\cite{matplotlib}, NumPy\cite{2020NumPy}, SciPy\cite{2020SciPy}, statsmodels\cite{statsmodels}, ray\cite{ray-2017}, Cython\cite{cython}, SAOImage DS9\cite{2003ASPC..295..489J}}

\subsection* {Data, Materials, and Code Availability} 

\vspace{2ex}
The Julia language SIRS package is freely available for download from the NASA GitHub. It includes several  jupyter notebook examples and an example showing how to apply SIRS reference correction using python-3 given the SIRS calibration file.


\bibliography{article}   
\bibliographystyle{spiejour}   


\vspace{2ex}\noindent\textbf{Bernard J. Rauscher} is an experimental astrophysicist at NASA's Goddard Space Flight Center. His research interests include astronomy instrumentation, extragalactic astronomy and cosmology, and the search for life on other worlds. Rauscher's work developing detector systems for JWST was recognized by a (shared) Congressional Space Act award and NASA's Exceptional Achievement Medal.

\vspace{2ex}\noindent\textbf{Dale J. Fixsen} is an experimental astrophysicist at the University of Maryland and NASA's Goddard Space Flight Center. He received his Ph.D. in physics from Princeton University in 1982. Fixsen's work has included developing mathematical models for COBE FIRAS, far infrared instrumentation, and building NIR astronomy instruments using  H1RGs.

\vspace{2ex}\noindent\textbf{Gregory Mosby, Jr.} is a Research Astrophysicist and detector scientist at NASA Goddard Space Flight Center. He received his BS degree in astronomy and physics from Yale University in 2009, and his MS and PhD degrees in astronomy from the University of Wisconsin - Madison in 2011 and 2016, respectively. His current research interests include near infrared detectors, astronomical instrumentation, and applications of machine learning to observational astronomy. He is a member of SPIE.


\listoffigures
\listoftables

\end{document}